%% file: Text-II.tex
\documentclass[11pt,twoside]{article}

\input{packages.tex}

\usepackage{graphicx}

\title{\huge
  Energetic Stability of the Solutions \\
  of the Einstein Field Equations for \\
  Spherically Symmetric Liquid Shells}

\author{\Large
  Jorge L. deLyra\footnote{Email: delyra@lmail.if.usp.br} \\[2ex]
  Universidade de São Paulo \\
  Instituto de Física \\
  Rua do Matão, 1371, \\
  05508-090 São Paulo, SP, Brazil}

\date{March 18, 2021}


\input{math-defs.tex}

\begin{document}\maketitle

\begin{abstract}
  \noindent
  We interpret the exact solutions previously obtained for spherically
  symmetric shells of liquid fluid in General Relativity in terms of the
  energies involved. In order to do this we make a change of variables in
  the field equations in order to introduce some integral expressions that
  are related to various parts of the energy. We then use these integrals
  in order to show that a certain parameter with dimensions of length,
  that was necessarily introduced into the solutions by the interface
  boundary conditions, is related to the binding energies of the
  gravitational systems.

  In sequence, we use this representation of the gravitational binding
  energy in order to discuss the energetic stability of the new solutions
  found. We include in the stability discussion the well-known interior
  Schwarzschild solution for a liquid sphere, which can be obtained as a
  specific limit of the solutions that were previously obtained for the
  liquid shells. We show that this particular family of solutions turns
  out to have zero binding energy and therefore to be a maximally unstable
  one, from the energetic point of view discussed here.

  We also perform a numerical exploration of the energetic stability
  criterion of the liquid shell solutions, all of which have strictly
  positive binding energies, and show that indeed there is a particular
  subset of the solutions which are energetically stable. All these
  solutions have the form of shells with non-vanishing internal radii.
  This reduces the original three-parameter family of liquid shell
  solutions to a two-parameter family of energetically stable solutions.
\end{abstract}

\newpage

\section{Introduction}\label{Sec01}

The issue of the energy in General Relativity is a difficult one, and its
discussion in specific examples quite often becomes involved and obscure.
The difficulties start at the very foundations of the theory, with the
impossibility of defining an energy-momentum tensor density for the
gravitational field itself, a problem which apparently is related to the
impossibility of localizing the energy of the gravitational field in the
general case~\cite{DiracGravity}.

However, a recently discovered new class of static and time-independent
exact solutions~\cite{LiquidShells} provides us with an opportunity to
discuss the subject in a clear, precise and complete manner. It leads to a
simple and clear characterization of all the energies involved in this
class of solutions, as well as a characterization of the relations among
them, which establishes an important connection with the fundamental
concept of the conservation of energy.

It is noteworthy that results similar to the ones we presented
in~\cite{LiquidShells} were obtained for the case of neutron stars, with
the Chandrasekhar equation of state~\cite{WeinbergGC}, by
Ni~\cite{NiNeutrStars} and Neslu\v{s}an~\cite{NeslusanNeutrStars}. Just as
in~\cite{LiquidShells}, the analysis of that case also led to an inner
vacuum region containing a singularity at the origin and a gravitational
field which is repulsive with respect to that origin. This tends to
indicate that these results are general at least to some extent. It is to
be expected that the ideas regarding the energy that we present here will
be useful in that case as well.

This paper is organized as follows: in the remainder of this introduction
we quickly review the new class of static and time-independent exact
solutions for liquid shells, as well as the interior Schwarzschild
solution, which can obtained from the new shell solutions in a certain
limit; in Section~\ref{Sec02} we establish certain general integral
formulas for all the energies involved; in Section~\ref{Sec03} we
establish the general physical interpretation of the energies involved,
including for both the shell solutions and the interior Schwarzschild
solution; in Section~\ref{Sec04} we perform a small numerical exploration
of the energetic stability of the shell solutions, and in
Section~\ref{Sec05} we state our conclusions.

\subsection{The Liquid Shell Solutions}\label{SSec1.1}

In a previous paper~\cite{LiquidShells} we established the solution of the
Einstein field equations for the case of a spherically symmetric shell of
liquid fluid located between the radial positions $r_{1}$ and $r_{2}$ of
the Schwarzschild system of coordinates. This is a three-parameter family
of solutions, which can be taken as any three of the four parameters
$r_{1}$, $r_{2}$, $M$ and $\rho_{0}$. The matter distribution is
characterized by the radii $r_{1}$ and $r_{2}$, by its total asymptotic
gravitational mass $M$, associated to the Schwarzschild radius $r_{M}$,
and by a matter energy density $\rho_{0}$ which is constant with the
radial Schwarzschild coordinate $r$ within $(r_{1},r_{2})$, and zero
outside that interval. In this work we will use the time-like signature
$(+,-,-,-)$, following~\cite{DiracGravity}. In terms of the coefficients
of the metric, for an invariant interval given in terms of the
Schwarzschild coordinates $(t,r,\theta,\phi)$ by

\begin{equation}\label{Eqn01}
  ds^{2}
  =
  \e{2\nu(r)}
  c^{2}dt^{2}
  -
  \e{2\lambda(r)}
  dr^{2}
  -
  r^{2}
  \left[
    d\theta^{2}
    +
    \sin^{2}(\theta)
    d\phi^{2}
  \right],
\end{equation}

\noindent
where $\exp[\nu(r)]$ and $\exp[\lambda(r)]$ are two positive functions of
only $r$, as was explained in~\cite{LiquidShells} the Einstein field
equations reduce to the set of three first-order differential equations

\noindent
\begin{eqnarray}
  \label{Eqn02}
  \left\{\rule{0em}{2.5ex}1-2\left[r\lambda'(r)\right]\right\}
  \e{-2\lambda(r)}
  & = &
        1-\kappa r^{2}\rho(r),
  \\
  \label{Eqn03}
  \left\{\rule{0em}{2.5ex}1+2\left[r\nu'(r)\right]\right\}
  \e{-2\lambda(r)}
  & = &
        1+\kappa r^{2}P(r),
  \\
  \label{Eqn04}
  \left[\rho(r)+P(r)\right]
  \nu'(r)
  & = &
        -P'(r),
\end{eqnarray}

\noindent
where $\rho(r)$ is the energy density of the matter, $P(r)$ is the
isotropic pressure, $\kappa=8\pi G/c^{4}$, $G$ is the universal
gravitational constant and $c$ is the speed of light. In these equations
the primes indicate differentiation with respect to $r$. Given these
equations, as presented in~\cite{LiquidShells} the complete solution for
$\lambda(r)$ is given by

\noindent
\begin{eqnarray}\label{Eqn05}
  \lambda(r)
  & = &
        \left\{
        \begin{array}{lcl}
          -\,
          \FFrac{1}{2}\,
          \ln\!\left(\FFrac{r+r_{\mu}}{r}\right)
          &
            \mbox{for}
          &
            0\;\leq r\leq r_{1},
          \\[3ex]
          -\,
          \FFrac{1}{2}\,
          \ln\!
          \left[
          \FFrac
          {
          \kappa\rho_{0}
          \left(r_{2}^{3}-r^{3}\right)
          +
          3\left(r-r_{M}\right)
          }
          {
          3r
          }
          \right]
          &
            \mbox{for}
          &
            r_{1}\leq r\leq r_{2},
          \\[3ex]
          -\,
          \FFrac{1}{2}\,
          \ln\!\left(\FFrac{r-r_{M}}{r}\right)
          &
            \mbox{for}
          &
            r_{2}\leq r<\infty,
        \end{array}
            \right.
\end{eqnarray}

\noindent
where $r_{M}=2GM/c^{2}$, while for $\nu(r)$ we have

\noindent
\begin{eqnarray}\label{Eqn06}
  \nu(r)
  & = &
        \left\{
        \begin{array}{lcl}
          \FFrac{1}{2}\,
          \ln\!
          \left(
          \FFrac{1-r_{M}/r_{2}}{1+r_{\mu}/r_{1}}
          \right)
          +
          \FFrac{1}{2}\,
          \ln\!\left(\FFrac{r+r_{\mu}}{r}\right)
          &
            \mbox{for}
          &
            0\;\leq r\leq r_{1},
          \\[3ex]
          \FFrac{1}{2}\,
          \ln\!
          \left(
          \FFrac{r_{2}-r_{M}}{r_{2}}
          \right)
          +
          \ln\!\left[z(r)\right]
          &
            \mbox{for}
          &
            r_{1}\leq r\leq r_{2},
          \\[3ex]
          \FFrac{1}{2}\,
          \ln\!\left(\FFrac{r-r_{M}}{r}\right)
          &
            \mbox{for}
          &
            r_{2}\leq r<\infty,
        \end{array}
            \right.
\end{eqnarray}

\noindent
and finally the pressure within the shell, that is, for
$r_{1}\leq r\leq r_{2}$, is given by

\begin{equation}\label{Eqn07}
  P(r)
  =
  \rho_{0}\,
  \frac{1-z(r)}{z(r)}.
\end{equation}

\noindent
This solution is valid under the condition that $r_{2}>r_{M}$. In all
these expressions we have that $r_{\mu}$ is given in terms of the
parameters characterizing the system by

\begin{equation}\label{Eqn08}
  r_{\mu}
  =
  \frac{\kappa\rho_{0}}{3}
  \left(
    r_{2}^{3}
    -
    r_{1}^{3}
  \right)
  -
  r_{M},
\end{equation}

\noindent
we have that $\rho_{0}$ is determined algebraically in terms of $r_{1}$,
$r_{2}$ and $r_{M}$ as the solution of the transcendental algebraic
equation

\noindent
\begin{eqnarray}\label{Eqn09}
  \sqrt{\frac{r_{2}}{3\left(r_{2}-r_{M}\right)}}
  & = &
        \sqrt
        {
        \frac
        {r_{1}}
        {
        \kappa\rho_{0}\left(r_{2}^{3}-r_{1}^{3}\right)
        +
        3\left(r_{1}-r_{M}\right)
        }
        }
        +
        \nonumber
  \\
  &   &
        +
        \frac{3}{2}
        \int_{r_{1}}^{r_{2}}dr\,
        \frac
        {\kappa\rho_{0}\;r^{5/2}}
        {
        \left[
        \kappa\rho_{0}\left(r_{2}^{3}-r^{3}\right)
        +
        3\left(r-r_{M}\right)
        \right]^{3/2}
        },
\end{eqnarray}

\noindent
and we have that the real function $z(r)$ is determined in terms of a
non-trivial elliptic real integral by the relation

\noindent
\begin{eqnarray}\label{Eqn10}
  z(r)
  & = &
        \sqrt
        {
        \frac
        {\kappa\rho_{0}\left(r_{2}^{3}-r^{3}\right)+3\left(r-r_{M}\right)}
        {r}
        }
        \times
        \nonumber
  \\
  &   &
        \times
        \left\{
        \sqrt{\frac{r_{2}}{3\left(r_{2}-r_{M}\right)}}
        +
        \frac{3}{2}
        \int_{r_{2}}^{r}ds\,
        \frac
        {\kappa\rho_{0}\;s^{5/2}}
        {
        \left[\kappa\rho_{0}\left(r_{2}^{3}-s^{3}\right)
        +
        3\left(s-r_{M}\right)\right]^{3/2}
        }
        \right\}.
\end{eqnarray}

\noindent
The relation shown in Equation~(\ref{Eqn08}) is a direct consequence of
the field equations and of the interface boundary conditions associated to
them. In~\cite{LiquidShells} we proved that, so long as the pressure of
the liquid is positive, we must have $r_{\mu}>0$. In fact, the hypotheses
of that proof can be weakened to require only that the pressure be
strictly positive at a single point. This strictly positive value of
$r_{\mu}$ implies that the solution has a singularity at the origin.
However, that singularity is not associated to an infinite concentration
of matter, but rather, as explained in~\cite{LiquidShells}, to zero energy
density at that point. Also, the solution introduces into the system the
new physical parameter $r_{\mu}$ with dimensions of length, which can be
associated to a mass parameter $\mu$ in the same way that $M$ is
associated to $r_{M}$.

\subsection{The Interior Schwarzschild Solution}\label{SSec1.2}

It is an interesting and somewhat remarkable fact that the well-known
interior Schwarzschild solution~\cite{SchwarzschildInternal,Wald} can be
obtained from our solution for a shell, even though the interior
Schwarzschild solution has no singularity at the origin, while our
solution always has that singularity. Curiously enough, we must start by
assuming that $r_{\mu}=0$, even though we proved in~\cite{LiquidShells}
that one must have $r_{\mu}>0$ in the shell solutions. The subtle point
here is that the proof given in~\cite{LiquidShells} relies on the
existence of a shell with $r_{1}>0$, while in the case of the interior
Schwarzschild solution we will have to use $r_{1}=0$, so that the shell
becomes a filled sphere. If we start by first putting $r_{\mu}=0$ and then
making $r_{1}\to 0$ in Equation~(\ref{Eqn08}), we are led to the relation

\begin{equation}\label{Eqn11}
  \kappa\rho_{0}
  =
  \frac{3r_{M}}{r_{2}^{3}},
\end{equation}

\noindent
so that we may substitute $\kappa\rho_{0}$ in terms of $r_{M}$ and the
radius $r_{2}$ of the resulting sphere. Following the usual notation for
the interior Schwarzschild solution, we now define a parameter $R$, with
dimensions of length, such that $R^{2}=r_{2}^{3}/r_{M}$, in terms of which
we have

\begin{equation}\label{Eqn12}
  \kappa\rho_{0}
  =
  \frac{3}{R^{2}}.
\end{equation}

\noindent
Note that the required condition that $r_{2}>r_{M}$ is translated here as
the condition that $R>r_{2}$. Making this substitution we have for
$\lambda(r)$ inside the resulting sphere, directly from the line in
Equation~(\ref{Eqn05}) for the case of the matter region, in the case in
which $r_{\mu}=0$ and $r_{1}\to 0$,

\begin{equation}\label{Eqn13}
  \lambda_{i}(r)
  =
  -\,
  \frac{1}{2}\,
  \ln\!\left[1-\left(\frac{r}{R}\right)^{2}\right],
\end{equation}

\noindent
which implies that for the radial metric coefficient we have

\begin{equation}\label{Eqn14}
  \e{-\lambda_{i}(r)}
  =
  \sqrt{1-\left(\frac{r}{R}\right)^{2}}.
\end{equation}

\noindent
In order to obtain $\nu(r)$ inside the sphere we must first work out the
function $z(r)$. Making the substitution of $\kappa\rho_{0}$ in terms of
$R$ in the result for $z(r)$ given in Equation~(\ref{Eqn10}) we get

\begin{equation}\label{Eqn15}
  z(r)
  =
  \sqrt{1-\left(\frac{r}{R}\right)^{2}}
  \left[
    \sqrt{\frac{r_{2}}{r_{2}-r_{M}}}
    +
    \frac{3}{2}
    \int_{r_{2}}^{r}ds\,
    \frac
    {s/R^{2}}
    {\left(1-s^{2}/R^{2}\right)^{3/2}}
  \right].
\end{equation}

\noindent
Is is now easy to see that in this case the remaining integral can be
done, and we get

\begin{equation}\label{Eqn16}
  z(r)
  =
  \frac{3}{2}
  -
  \frac{1}{2}\,
  \sqrt{\frac{r_{2}}{r_{2}-r_{M}}}
  \sqrt{1-\left(\frac{r}{R}\right)^{2}}.
\end{equation}

\noindent
Using again the definition of $R$, which implies that we have
$r_{M}/r_{2}=\left(r_{2}/R\right)^{2}$, we may write this as

\begin{equation}\label{Eqn17}
  z(r)
  =
  \frac{3}{2}
  -
  \frac{1}{2}\,
  \sqrt{\frac{1-\left(r/R\right)^{2}}{1-\left(r_{2}/R\right)^{2}}}.
\end{equation}

\noindent
Note that we have $z(r_{2})=1$, which corresponds to $P(r_{2})=0$, so that
the boundary conditions for $z(r)$ and $P(r)$ at $r_{2}$ are still
satisfied. From this we may now obtain all the remaining results for the
interior Schwarzschild solution. From the line in Equation~(\ref{Eqn06})
for the case of the matter region, in the case in which $r_{\mu}=0$ and
$r_{1}\to 0$, we get for $\nu(r)$ in the interior of the sphere

\begin{equation}\label{Eqn18}
  \nu_{i}(r)
  =
  \frac{1}{2}\,
  \ln\!\left[1-\left(\frac{r_{2}}{R}\right)^{2}\right]
  +
  \ln\!
  \left[
    \frac{3}{2}
    -
    \frac{1}{2}\,
    \sqrt{\frac{1-\left(r/R\right)^{2}}{1-\left(r_{2}/R\right)^{2}}}
  \right],
\end{equation}

\noindent
which implies that for the temporal metric coefficient we have

\begin{equation}\label{Eqn19}
  \e{\nu_{i}(r)}
  =
  \frac{3}{2}
  \sqrt{1-\left(\frac{r_{2}}{R}\right)^{2}}
  -
  \frac{1}{2}\,
  \sqrt{1-\left(\frac{r}{R}\right)^{2}}.
\end{equation}

\noindent
Finally, from Equation~(\ref{Eqn07}), in the case in which $r_{\mu}=0$ and
$r_{1}\to 0$, we get for the pressure $P(r)$ within the sphere

\begin{equation}\label{Eqn20}
  P(r)
  =
  \rho_{0}\,
  \frac
  {\sqrt{1-\left(r/R\right)^{2}}-\sqrt{1-\left(r_{2}/R\right)^{2}}}
  {3\sqrt{1-\left(r_{2}/R\right)^{2}}-\sqrt{1-\left(r/R\right)^{2}}}.
\end{equation}

\noindent
These are indeed the correct results for the case of the interior
Schwarzschild solution. Note that all the arguments of the logarithms and
of the square roots are positive due to the conditions that $R>r_{2}>r$.
Note also that in the $r_{1}\to 0$ limit the lines in
Equations~(\ref{Eqn05}) and~(\ref{Eqn06}) for the case of the inner vacuum
region become irrelevant, since this region reduces to a single point. On
the other hand, the lines for the case of the outer vacuum region do not
change at all.

It is therefore apparent that the $r_{1}\to 0$ limit of our shell
solutions does reproduce the interior Schwarzschild solution, so long as
we adopt the value zero for $r_{\mu}$. Our interpretation of these facts
is that the $r_{1}\to 0$ limit to the interior Schwarzschild solution is a
{\em non-uniform} one, in which we have to leave out one point, the
origin. In the $r_{1}\to 0$ limit the singularity of the shell solutions
becomes a strictly point-like one, and therefore a removable one, by a
simple continuity criterion. This is certainly the case for the energy
density $\rho(r)$, which in the limit is non-zero everywhere around the
origin but at a single point, the origin itself. The same is true for the
pressure $P(r)$, which in the limit is also non-zero around the origin but
at the origin itself. Similar situations hold for $\lambda(r)$ and
$\nu(r)$, as is not difficult to see numerically. It seems that all these
functions converge in the $r_{1}\to 0$ limit to functions with a
point-like removable discontinuity at the origin.

\section{Integral Expressions for the Energies}\label{Sec02}

It is possible to express the masses $M$ and $\mu$, as well as the
corresponding energies $Mc^{2}$ and $\mu c^{2}$, which are associated to
the parameters with dimensions of length $r_{M}=2MG/c^{2}$ and
$r_{\mu}=2\mu G/c^{2}$ that appear in the exact solutions described in
Section~\ref{Sec01}, as integrals of the matter energy density $\rho(r)$
over coordinate volumes, in a way similar to what is usually done for $M$
in the literature~\cite{MisnerThorneWheeler,WeinbergGC}, but leading to
very different results in the case of the shell solutions. In order to do
this in a simple and organized way, we first change variables in the field
equations from $\lambda(r)$ to $\beta(r)$, which is defined to be such
that

\begin{equation}\label{Eqn21}
  \e{2\lambda(r)}
  =
  \frac{r}{r-r_{M}\beta(r)},
\end{equation}

\noindent
which then implies that we have for the corresponding derivatives

\begin{equation}\label{Eqn22}
  2r\lambda'(r)
  =
  -r_{M}\,
  \frac{\beta(r)-r\beta'(r)}{r-r_{M}\beta(r)}.
\end{equation}

\noindent
Note that $\beta(r)=0$ corresponds to $\lambda(r)=0$ and therefore to
$\exp[2\lambda(r)]=1$ for the radial coefficient of the metric. In such
cases the variations of the radial coordinate are equal to the variations
of the corresponding proper lengths. Substituting these expressions in the
component field equation shown in Equation~(\ref{Eqn02}) a very simple
relation giving the derivative of $\beta(r)$ in terms of $\rho(r)$
results,

\begin{equation}\label{Eqn23}
  \beta'(r)
  =
  \frac{\kappa r^{2}\rho(r)}{r_{M}}.
\end{equation}

\noindent
Therefore, wherever $\rho(r)=0$, we have that $\beta(r)$ is a constant.
Note that these facts are completely general for the spherically symmetric
static case, in the sense that they are not limited to the case in which
$\rho(r)$ is constant within the matter region. It then follows from
Equation~(\ref{Eqn05}) that we have that $\beta(r)=1>0$ in the outer
vacuum region, and in particular at $r_{2}$, and that we have that
$\beta(r)=-r_{\mu}/r_{M}<0$ in the inner vacuum region, and in particular
at $r_{1}$. Since $\beta(r)$ is a continuous function that goes from
negative values at $r_{1}$ to positive values at $r_{2}$, it follows that
there is a radial position $r_{z}$ within the matter region where
$\beta(r_{z})=0$, regardless of whether or not $\rho(r)$ is constant
within the shell. At this particular radial position we also have that
$\lambda(r_{z})=0$.

Let us now consider the integral of the energy density over a coordinate
volume within the matter region, where $\rho(r)\neq 0$, say from an
arbitrary point $r_{a}$ to another point $r_{b}>r_{a}$,

\begin{equation}\label{Eqn24}
  \int_{r_{a}}^{r_{b}}dr
  \int_{0}^{\pi}d\theta
  \int_{0}^{2\pi}d\phi\,
  r^{2}\sin(\theta)\rho(r)
  =
  4\pi
  \int_{r_{a}}^{r_{b}}dr\,
  r^{2}\rho(r),
\end{equation}

\noindent
where we integrated over the angles. Note that this is not an integral
over the proper volume, but just an integral over the coordinate volume,
since we are missing here the remaining factor $\exp[\lambda(r)+\nu(r)]$
of the Jacobian $\sqrt{-g}$. Since we have the three special points
$r_{1}$, $r_{z}$ and $r_{2}$ where the values of $\beta(r)$ are known, let
us consider now the integral of the energy density over the coordinate
volume from $r_{z}$ to $r_{2}$. Using Equation~(\ref{Eqn23}) we get

\begin{equation}\label{Eqn25}
  4\pi
  \int_{r_{z}}^{r_{2}}dr\,
  r^{2}\rho(r)
  =
  4\pi
  \frac{r_{M}}{\kappa}
  \int_{r_{z}}^{r_{2}}dr\,
  \beta'(r).
\end{equation}

\noindent
One can now see that the integral is trivial, and since we have that
$\beta(r_{z})=0$ and that $\beta(r_{2})=1$, we get

\begin{equation}\label{Eqn26}
  Mc^{2}
  =
  4\pi
  \int_{r_{z}}^{r_{2}}dr\,
  r^{2}\rho(r),
\end{equation}

\noindent
where we have replaced $\kappa$ and $r_{M}$ by their values in terms of
$M$ and $c$. We have therefore an expression for the energy $Mc^{2}$ in
terms of a coordinate volume integral of the energy density. Note however
that the integral does not run over the whole matter region, since it
starts at $r_{z}$ rather than at $r_{1}$. In a similar way, if we consider
the integral from $r_{1}$ to $r_{z}$, we get

\begin{equation}\label{Eqn27}
  4\pi
  \int_{r_{1}}^{r_{z}}dr\,
  r^{2}\rho(r)
  =
  4\pi
  \frac{r_{M}}{\kappa}
  \int_{r_{1}}^{r_{z}}dr\,
  \beta'(r).
\end{equation}

\noindent
Once again one can see that the integral is trivial, and since we have
that $\beta(r_{z})=0$ and that $\beta(r_{1})=-r_{\mu}/r_{M}$, we now get

\begin{equation}\label{Eqn28}
  \mu c^{2}
  =
  4\pi
  \int_{r_{1}}^{r_{z}}dr\,
  r^{2}\rho(r),
\end{equation}

\noindent
where we have replaced $\kappa$ and $r_{\mu}$ by their values in terms of
$\mu$ and $c$. We have therefore an expression for the energy $\mu c^{2}$
in terms of a coordinate volume integral of the energy density.

If we now consider the integral over the whole matter region, due to the
additive property of the integrals over the union of disjoint domains,
using Equations~(\ref{Eqn26}) and~(\ref{Eqn28}) we obtain the result that

\begin{equation}\label{Eqn29}
  4\pi
  \int_{r_{1}}^{r_{2}}dr\,
  r^{2}\rho(r)
  =
  \mu c^{2}
  +
  Mc^{2}.
\end{equation}

\noindent
This is a sum of energies, and is therefore also an energy, to which we
will associate a mass parameter $M_{u}$, such that this energy is given by
$M_{u}c^{2}$, so that we have the relation

\begin{equation}\label{Eqn30}
  M_{u}c^{2}
  =
  \mu c^{2}
  +
  Mc^{2}.
\end{equation}

\noindent
We see therefore that the point $r_{z}$ where $\beta(r_{z})=0$ and
therefore $\lambda(r_{z})=0$ plays a particular role when it comes to the
determination of the energies involved.

Note that all this is true for any function $\rho(r)$ within the matter
region. For our specific case here, with a constant $\rho_{0}$, we find
from Equation~(\ref{Eqn05}) that we have within the matter region

\begin{equation}\label{Eqn31}
  \beta(r)
  =
  1-\frac{\kappa\rho_{0}}{3r_{M}}\,\left(r_{2}^{3}-r^{3}\right),
\end{equation}

\noindent
so that in this case we have for the zero $r_{z}$ of $\beta(r)$

\begin{equation}\label{Eqn32}
  r_{z}
  =
  \left(r_{2}^{3}-\frac{3r_{M}}{\kappa\rho_{0}}\right)^{1/3}.
\end{equation}

\noindent
Note that, although all these integrals are written in terms of the energy
density $\rho(r)$ of the matter, none of them represents just the energy
of only the matter itself. In fact we must now interpret the meaning of
each one of these expressions, which is what we will do in the next
section.

\section{Physical Interpretation of the Energies }\label{Sec03}

Of the three energies at play here, namely $M_{u}c^{2}$, $\mu c^{2}$ and
$Mc^{2}$, only the last one has a well established meaning at this point.
Since $M$ is the asymptotic gravitational mass of the system, that is, the
gravitational mass seen as the source of the gravitational field at large
radial distances, the standard interpretation in General Relativity is
that the energy $Mc^{2}$ is the total energy of this gravitational system,
bound into the shell by the gravitational interactions, and which from now
on we will simply call the {\em bound system}. It includes both the energy
of the matter in the bound state and the energy stored in the
gravitational field itself, also in this bound state. The energy density
$\rho(r)$ is the amount of energy of the matter, per unit volume, as seen
by a stationary local observer at the radial position $r$.

Our first task here is to establish the physical interpretation of the
energy $M_{u}c^{2}$. In order to do this, the first thing to be done is to
define an {\em unbound system} related to our bound system as defined
above. This unbound system is what we get when we scatter all the elements
of the shell to very large distances from each other, in order to
eliminate all the gravitational interactions, but without making any
changes in the energy content of the matter. We will show here that the
energy $M_{u}c^{2}$ is the total energy of this unbound system. We will do
this by performing a mathematical transformation on the integral in
Equation~(\ref{Eqn29}), which with the use of Equation~(\ref{Eqn30}) leads
to the following expression in terms of a volume integral

\begin{equation}\label{Eqn33}
  M_{u}c^{2}
  =
  \int_{r_{1}}^{r_{2}}dr
  \int_{0}^{\pi}d\theta
  \int_{0}^{2\pi}d\phi\,
  r^{2}\sin(\theta)\rho(r).
\end{equation}

\noindent
The transformation, applied to the right-hand side of this equation, will
allow us to interpret the meaning of the left-hand side. This will be done
in a general way, for any function $\rho(r)$ within the matter region.
This transformation will consist in fact of the construction of a second
integral, based on the concept of the Riemann sums of the volume integral
shown in Equation~(\ref{Eqn33}).

\begin{figure}[t]
  \centering
  {\color{white}\rule{\textwidth}{0.1ex}}
  %
  \epsfig{file=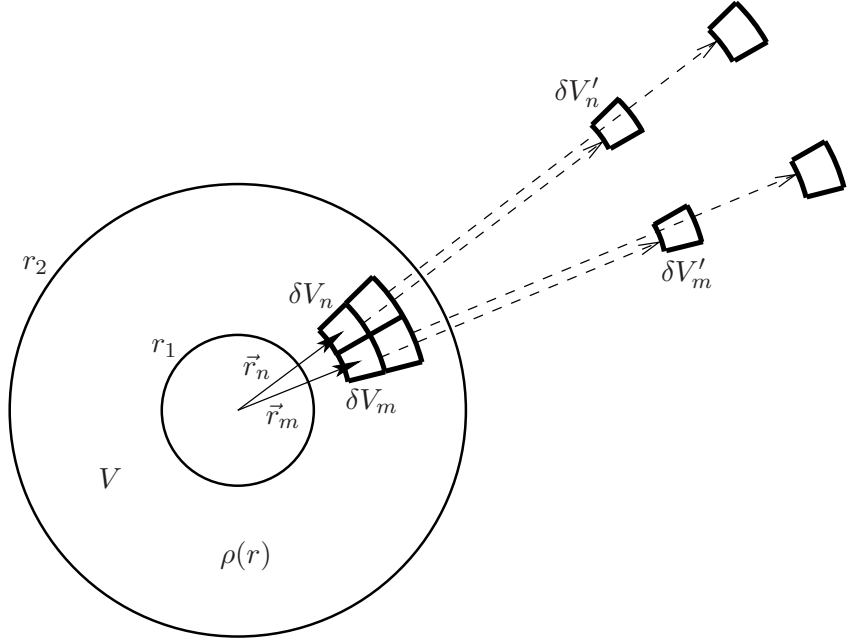,scale=1.0,angle=0}
  %
  \caption{Illustration of the geometrical transformation of the integral
    over the shell.}
  \label{Fig01}
\end{figure}

Let us consider therefore an arbitrary Riemann partition of the integral
in Equation~(\ref{Eqn33}), consisting of a finite number of cells
$\delta V_{n}$ with coordinate volume and linear coordinate dimensions
below certain maximum values, where $n\in\{1,\ldots,N\}$. By definition of
a partition the sum of all these volume elements is equal to the
coordinate volume $V$ of the shell,

\begin{equation}\label{Eqn34}
  V
  =
  \sum_{n=1}^{N}
  \delta V_{n},
\end{equation}

\noindent
where we will assume that each volume element is at the spatial position
$\vec{r}_{n}$, as illustrated in Figure~\ref{Fig01}. The energy
$M_{u}c^{2}$ can therefore be written as the integration limit of the
Riemann sum over this partition,

\begin{equation}\label{Eqn35}
  M_{u}c^{2}
  =
  \lim_{N\to\infty}
  \sum_{n=1}^{N}
  \rho(r_{n})\delta V_{n},
\end{equation}

\noindent
where $r_{n}=|\vec{r}_{n}|$. We now consider the mathematical
transformation in which we map each volume element $\delta V_{n}$ at
$\vec{r}_{n}$ onto an identical volume element $\delta V'_{n}$ at the
coordinate position $\vec{r}^{\;\prime}_{n}=\alpha\vec{r}_{n}$, for some
large positive real number $\alpha$, without changing the coordinate
volume of the volume elements. The result is a new set of volume elements,
all at large distances from each other, whose sum is still equal to the
coordinate volume of the shell,

\begin{equation}\label{Eqn36}
  V
  =
  \sum_{n=1}^{N}
  \delta V'_{n},
\end{equation}

\noindent
The geometrical transformation leading to the construction of the new
integral is illustrated in Figure~\ref{Fig01}. Note that no physical
transport of the matter or of the energy within the volume elements
$\delta V_{n}$ of the shell is meant here, so that there are no actual
physical transformations involved.

After defining the volume elements $\delta V'_{n}$ ta large distances in
this fashion, we now put within each one of these new volume elements
exactly the same amount of mass and energy that we have in the
corresponding coordinate volume elements $\delta V_{n}$ of the shell. This
means putting into each volume element $\delta V'_{n}$ at infinity the
same numbers of the same types of particles, as well as the same amount of
thermal energy and pressure, as seen by a stationary local observer at the
position $\vec{r}^{\;\prime}_{n}$, that a stationary local observer at
$\vec{r}_{n}$ sees within $\delta V_{n}$. Is other words, we associate to
each volume element at infinity the same value of the energy density
$\rho(r'_{n})=\rho(r_{n})$ that we had for the corresponding volume
element of the shell, where $r'_{n}=|\vec{r}^{\;\prime}_{n}|$ and
$r_{n}=|\vec{r}_{n}|$.

For large values of $\alpha$ these elements of mass and energy within
$\delta V'_{n}$ are all at large distances from each other, so as to
render the gravitational interactions among them negligible. In the
$\alpha\to\infty$ limit all the gravitational interactions among the
volume elements $\delta V'_{n}$ go to zero. Besides, in the integration
limit each element of mass and energy so constructed tends to zero, so
that the gravitational self-interactions within each volume element also
become negligible. However, independently of either limit, by construction
the total coordinate volume of the elements of volume at infinity remains
equal to the coordinate volume of the shell. Therefore, by construction
the corresponding sum of all the energy elements of energy at infinity is
the same as the Riemann sum that appears in Equation~(\ref{Eqn35}),

\begin{equation}\label{Eqn37}
  \sum_{n=1}^{N}
  \rho(r'_{n})\delta V'_{n}
  =
  \sum_{n=1}^{N}
  \rho(r_{n})\delta V_{n}.
\end{equation}

\noindent
Now, at radial infinity spacetime is flat, so that the coordinate volume
of each volume element $\delta V'_{n}$ coincides with its proper volume,
and hence the energy element $\rho(r'_{n})\delta V'_{n}$ is the total
energy of that element of matter, so that the sum of all these energy
elements is the total energy of the matter at infinity. In other words,
once we take the integration limit the integral given in
Equation~(\ref{Eqn33}) gives us the total energy of the system at
infinity, which is free from all gravitational bindings. Hence we will
name the quantity $M_{u}c^{2}$ the {\em total energy of the unbound
  system}. This is the total energy of the system when all gravitational
interactions have been eliminated by increasing without limit the
distances among its elements. This is in both analogy and contrast with
the quantity $Mc^{2}$, which is the {\em total energy of the bound
  system}, after all its parts have been brought together to form the
shell.

Note that this whole argument is general, in the sense that it is not
limited to the case in which $\rho(r)=\rho_{0}$ is a constant. In our case
here, since $\rho(r)=\rho_{0}$ is a constant, the total energy of the
unbound system is just the product of $\rho_{0}$ by the coordinate volume
$V$ of the shell,

\begin{equation}\label{Eqn38}
  M_{u}c^{2}
  =
  \rho_{0}V.
\end{equation}

\noindent
Our next task here is to establish the physical interpretation of the
energy $\mu c^{2}$. From Equation~(\ref{Eqn30}) we have that the energy
parameter $\mu c^{2}$ is the difference between the total energy of the
unbound system and the total energy of the bound system,

\begin{equation}\label{Eqn39}
  \mu c^{2}
  =
  M_{u}c^{2}
  -
  Mc^{2},
\end{equation}

\noindent
and therefore we conclude that it is the {\em binding energy} of the
system. It is the amount of energy that must be given to the system in
order to disperse its elements to infinity, thus eliminating all the
gravitational bindings between those elements. It is also the amount of
energy that must be dissipated by the system during the process of its
assembly into the bound system, stating from the unbound system at
infinity. The theorem we proved in~\cite{LiquidShells}, in the
$\rho(r)=\rho_{0}$ case that we have here, namely that we must have
$r_{\mu}>0$, is equivalent to the statement that the bound system must
have a finite, positive and non-zero binding energy. This is, of course,
closely related to the attractive nature of the gravitational interaction
between particles.

Note that, although all these integrals are written in terms of the energy
density $\rho(r)$ of the matter, the energy $Mc^{2}$ is {\em not} the
energy $M_{m}c^{2}$ of just the matter within the bound system. That would
be given by the integral with the full Jacobian factor $\sqrt{-g}$, where
$g$ is the determinant of $g_{\mu\nu}$, which in our case here results in

\begin{equation}\label{Eqn40}
  M_{m}c^{2}
  =
  4\pi
  \int_{r_{1}}^{r_{2}}dr\,
  r^{2}\e{\lambda(r)+\nu(r)}\rho(r).
\end{equation}

\noindent
As a partial consistency check, it is not difficult to verify that this
energy is always smaller than $M_{u}c^{2}$, due to the fact that the
exponent $\lambda(r)+\nu(r)$ is always negative within the matter
region. In order to show this we just take the difference between the
component field equations shown in Equations~(\ref{Eqn03})
and~(\ref{Eqn02}), thus obtaining

\begin{equation}\label{Eqn41}
  \left[\lambda(r)+\nu(r)\right]'
  =
  \frac{\kappa}{2}\,
  \e{2\lambda(r)}
  r\left[\rho(r)+P(r)\right].
\end{equation}

\noindent
Since all quantities appearing on the right-hand side are positive or
zero, we may conclude that the derivative of the exponent is non-negative.
However, we have that $\lambda(r_{2})+\nu(r_{2})=0$, since this exponent
is identically zero within the outer vacuum region. It follows that

\begin{equation}\label{Eqn42}
  \lambda(r)+\nu(r)
  <
  0,
\end{equation}

\noindent
and therefore that

\begin{equation}\label{Eqn43}
  \e{\lambda(r)+\nu(r)}
  <
  1,
\end{equation}

\noindent
throughout the whole matter region, with the exception of the single point
$r_{2}$ where the exponential is equal to one. Therefore, it follows for
the two integrals that

\begin{equation}\label{Eqn44}
  4\pi
  \int_{r_{1}}^{r_{2}}dr\,
  r^{2}\e{\lambda(r)+\nu(r)}\rho(r)
  <
  4\pi
  \int_{r_{1}}^{r_{2}}dr\,
  r^{2}\rho(r),
\end{equation}

\noindent
and therefore that $M_{m}c^{2}<M_{u}c^{2}$. The difference
$Mc^{2}-M_{m}c^{2}$ is the part of the energy of the bound system which is
not the energy of the matter itself, but rather the energy stored in the
gravitational field. In general, in order to determine this difference,
$M_{m}c^{2}$ has to be calculated numerically.

\subsection{Energetic Stability}\label{SSec3.1}

This interpretation of the parameters involved leads right away to the
idea that we may define a notion of {\em energetic stability} of the
solutions obtained, in the general spirit of the principle of virtual
work. Given certain constraints regarding some of the parameters of the
solutions, we may obtain the parameter $r_{\mu}$ as a function of the
remaining parameters of the system. Within this class of solutions, if
there are two with different values of $r_{\mu}$, which is proportional to
the binding energy $\mu c^{2}$, then in principle the constrained system
will tend to go from the one with the smaller value of $r_{\mu}$ to the
one with the larger value, given the existence of a permissible path
between the two solutions. This type of analysis allows us to acquire some
information about the dynamical behavior of the system, without having to
find explicitly the corresponding time-dependent solutions.

Let us exemplify this with our current system, in a way that is physically
illustrative. Our system contains four parameters, namely $r_{1}$,
$r_{2}$, $r_{M}$ and $\rho_{0}$, of which only three are independent. As
was explained in~\cite{LiquidShells}, these four parameters are related by
the condition in Equation~(\ref{Eqn09}). Given any three of the
parameters, that equation can be used to determine the fourth in terms of
those three. Let us assume that we are given fixed values of both $M$ and
$\rho_{0}$, thus determining the local properties of the matter and the
total amount of energy of the bound system. This is equivalent to fixing
$r_{M}$ and $\rho_{0}$, and therefore the result of solving
Equation~(\ref{Eqn09}) is to establish $r_{1}$ as a function of
$r_{2}$. We therefore are left with a collection of solutions parametrized
by a single real parameter, the external radius $r_{2}$. We may then
determine $r_{\mu}(r_{2})$ and verify whether this function has a single
local maximum at a certain value of $r_{2}$. This then identifies that
particular solution which is stable, or that has the largest binding
energy, among all others, given the constraints described.

Another approach, slightly more indirect, but perhaps simpler and more
physically compelling, would be to keep constant the local parameter
$\rho_{0}$ and the energy $M_{u}c^{2}$ of the unbound system. This fixes
the local properties of the matter and the total energy of the unbound
system that we start with, and we may then ask which is the solution that
corresponds to the most tightly bound system that can be assembled from
that unbound system. Since the energy of the unbound system is the product
of $\rho_{0}$ by the coordinate volume $V$ of the shell, as can be seen in
Equation~(\ref{Eqn38}), keeping fixed both $\rho_{0}$ and $M_{u}$
corresponds to keeping fixed at a value $V_{0}$ that coordinate volume,
which is given by

\begin{equation}\label{Eqn45}
  V_{0}
  =
  \frac{4\pi}{3}
  \left(
    r_{2}^{3}
    -
    r_{1}^{3}
  \right).
\end{equation}

\noindent
This immediately determines $r_{2}$ as a simple function $r_{2}(r_{1})$ of
$r_{1}$. Then solving Equation~(\ref{Eqn09}) results in $r_{M}$ being
given as a function $r_{M}(r_{1})$ of $r_{1}$ for the fixed value of
$\rho_{0}$ and the fixed coordinate volume $V_{0}$. This corresponds to
the energy of the bound system with internal radius $r_{1}$, for the given
fixed values of $\rho_{0}$ and $V_{0}$. The minimum of this function gives
us the value of $r_{1}$ that corresponds to the most tightly bound system
that can be assembled from a given unbound system. Other solutions in the
same family, with other values of $r_{1}$, will tend to decay into this
one, given a permissible decay path between the two solutions involved.
We will execute this program numerically in Section~\ref{Sec04}.

We saw that in the case of the interior Schwarzschild solution we have the
value zero for $r_{\mu}$. This implies that the resulting solution has
zero gravitational binding energy, and that its energy is the same as the
energy of the corresponding unbound system, which is a very strange and
even bizarre situation indeed. This means that the resulting solution is
not only energetically unstable, but that it is in fact {\em maximally}
energetically unstable, since the bound system cannot possibly have more
energy than the unbound system. Given a permissible path, in principle one
would be able to disassemble the matter distribution of the interior
Schwarzschild solution, taking every element of matter do infinity,
without giving any energy at all to the system. This is quite unrealistic,
and may be the reason why this solution has never proved to be a very
useful one.

\section{Numerical Exploration of the Binding Energy}\label{Sec04}

Here we will explore numerically the issues of the binding energy and of
the energetic stability of the shell solutions. In this exploration we
will keep fixed the local energy density parameter $\rho_{0}$, as well as
the total energy $M_{u}c^{2}$ of the unbound system. Our objective will be
then to determine the existence and the parameters of the maximally bound
shell solution. We will do this by calculating the energy $Mc^{2}$ of the
bound system and showing that it has a point of minimum as a function of
$r_{1}$. Since we keep fixed the parameter $\rho_{0}$, and since the
energy of the unbound system is given by $M_{u}c^{2}=\rho_{0}V_{0}$, this
implies that we also keep fixed the coordinate volume $V_{0}$ of the
shell, given in Equation~(\ref{Eqn45}), which immediately establishes
$r_{2}$ as a given function of $r_{1}$,

\begin{equation}\label{Eqn46}
  r_{2}(r_{1})
  =
  \left(r_{1}^{3}+\frac{3V_{0}}{4\pi}\right)^{1/3}.
\end{equation}

\noindent
Therefore, of the three free parameters of our solutions, which can be
taken to be $r_{1}$, $r_{2}$ and $\rho_{0}$, one is being kept fixed and
another is a given function, so that we are left with only one free
parameters, which we will take to be $r_{1}$. Under these circumstances we
have that $r_{M}$, and therefore both the mass $M$ and the energy $Mc^{2}$
of the bound system, are functions of $r_{1}$, with values that are left
to be determined numerically.

\begin{figure}[t]
  \centering
  {\color{white}\rule{\textwidth}{0.1ex}}
  %
  \epsfig{file=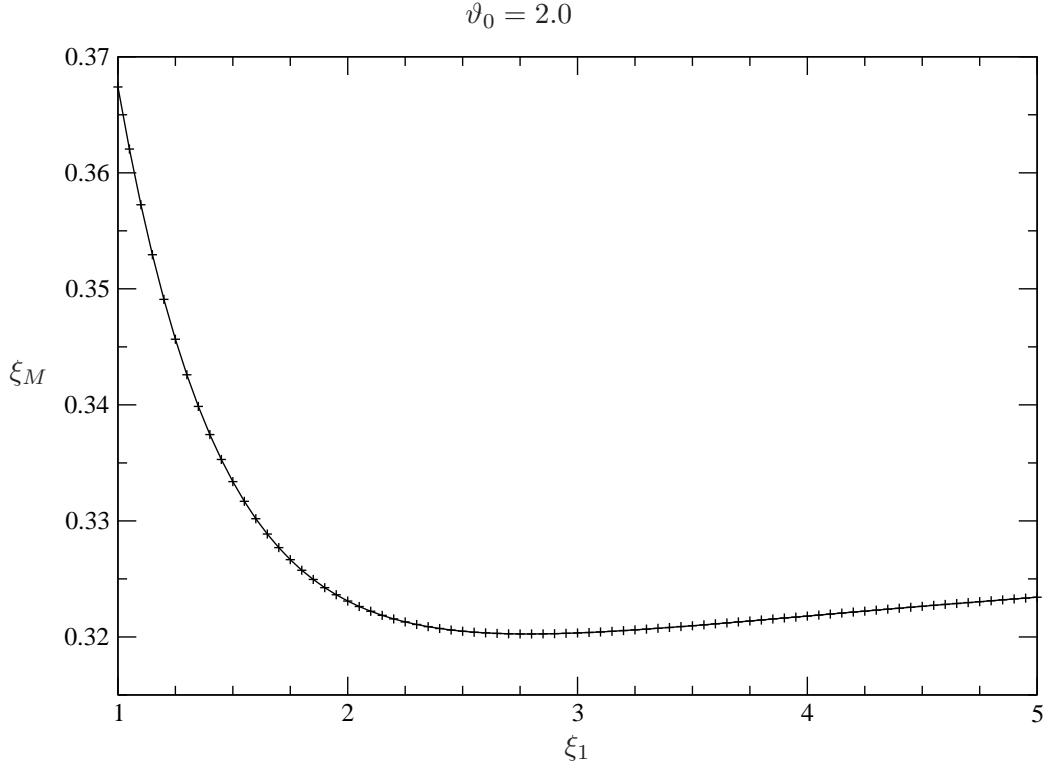,scale=1.0,angle=0}
  %
  \caption{Graph of the energy of the bound system as a function of
    $\xi_{1}$, for a fixed energy of the unbound system, given by
    $\vartheta_{0}=2$, and with $\xi_{1}$ in $[1,5]$.}
  \label{Fig02}
\end{figure}

\begin{figure}[t]
  \centering
  {\color{white}\rule{\textwidth}{0.1ex}}
  %
  \epsfig{file=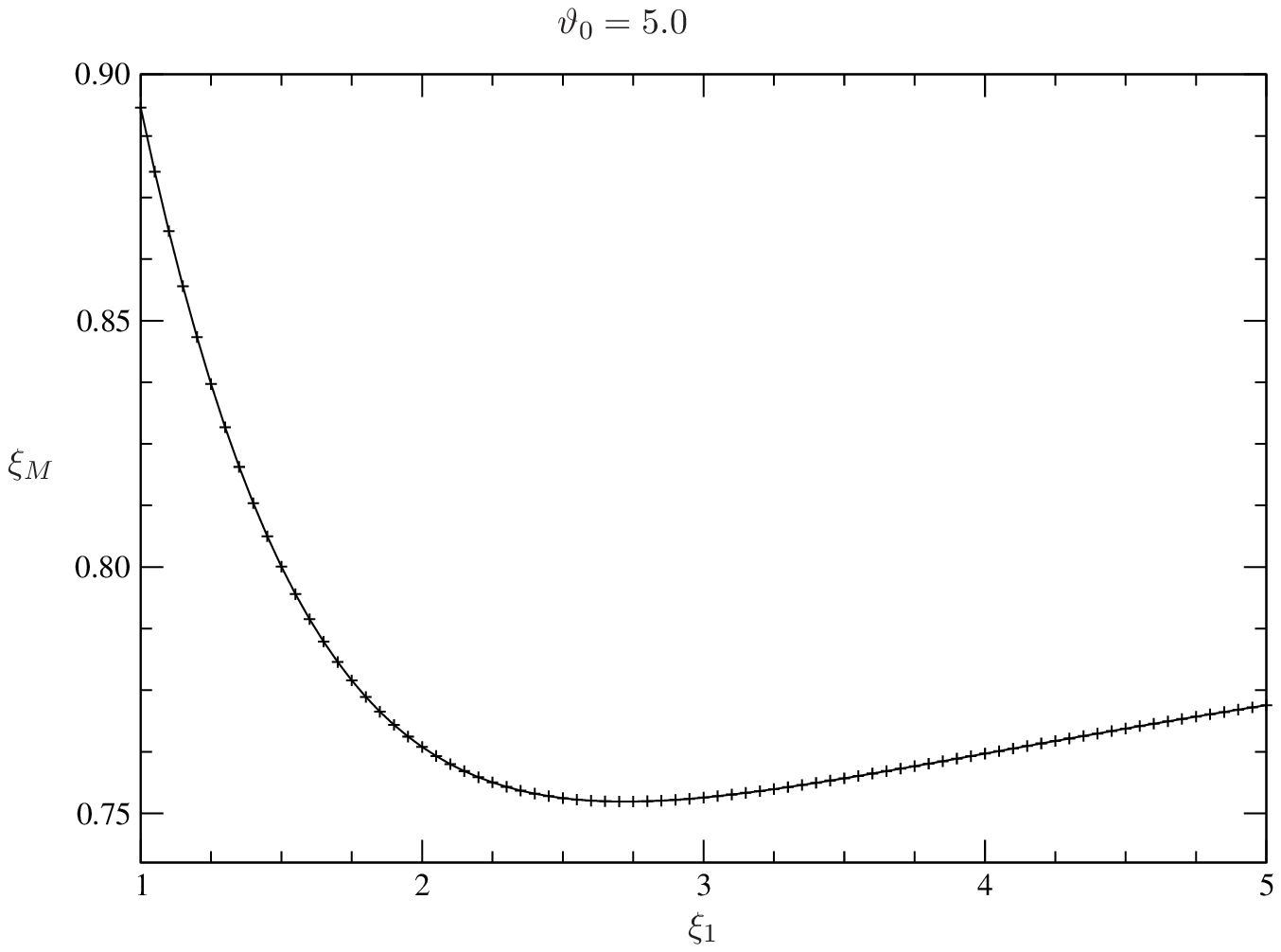,scale=1.0,angle=0}
  %
  \caption{Graph of the energy of the bound system as a function of
    $\xi_{1}$, for a fixed energy of the unbound system, given by
    $\vartheta_{0}=5$, and with $\xi_{1}$ in $[1,5]$.}
  \label{Fig03}
\end{figure}

\begin{figure}[t]
  \centering
  {\color{white}\rule{\textwidth}{0.1ex}}
  %
  \epsfig{file=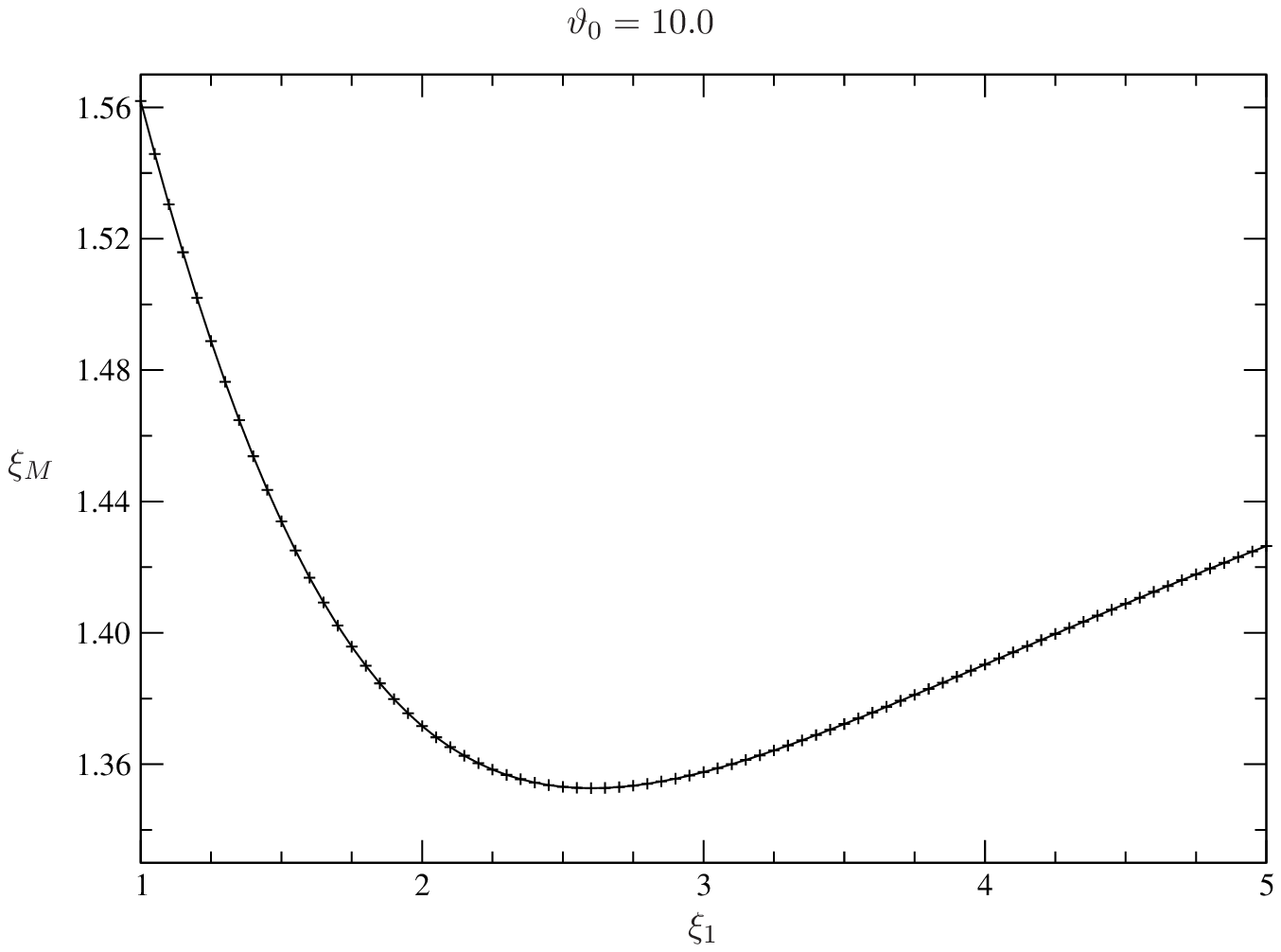,scale=1.0,angle=0}
  %
  \caption{Graph of the energy of the bound system as a function of
    $\xi_{1}$, for a fixed energy of the unbound system, given by
    $\vartheta_{0}=10$, and with $\xi_{1}$ in $[1,5]$.}
  \label{Fig04}
\end{figure}

\begin{figure}[t]
  \centering
  {\color{white}\rule{\textwidth}{0.1ex}}
  %
  \epsfig{file=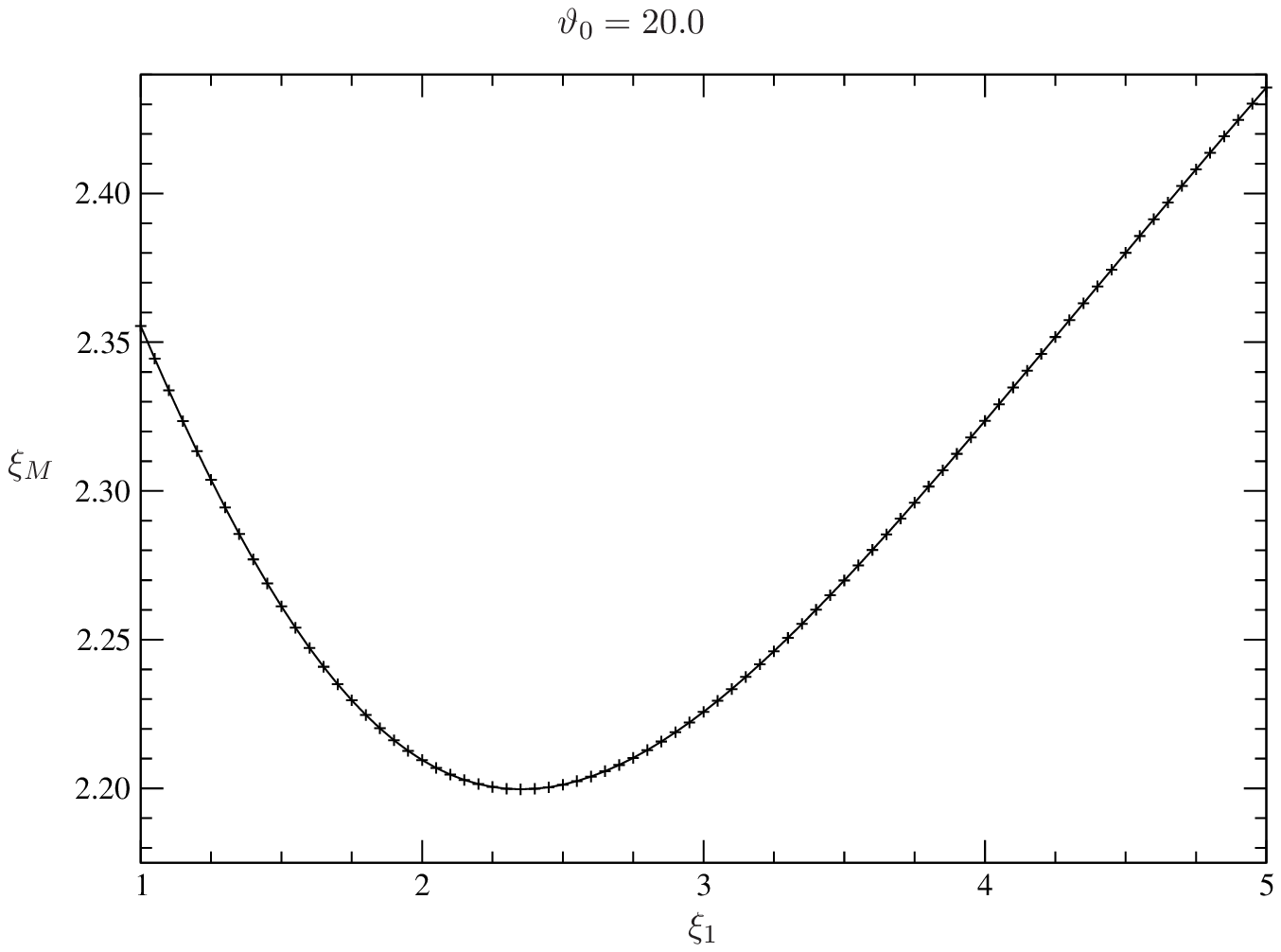,scale=1.0,angle=0}
  %
  \caption{Graph of the energy of the bound system as a function of
    $\xi_{1}$, for a fixed energy of the unbound system, given by
    $\vartheta_{0}=20$, and with $\xi_{1}$ in $[1,5]$.}
  \label{Fig05}
\end{figure}

In order to perform the numerical work it is convenient to first rescale
the variables, creating a set of equivalent dimensionless variables. Since
under these conditions $\kappa\rho_{0}$ is a constant which has dimensions
of inverse square length, we will define a constant $r_{0}$ with
dimensions of length by

\begin{equation}\label{Eqn47}
  r_{0}
  =
  \frac{1}{\sqrt{\kappa\rho_{0}}}.
\end{equation}

\noindent
Having now the known constant $r_{0}$, we use it in order to define the
set of dimensionless parameters given by

\noindent
\begin{eqnarray}\label{Eqn48}
  \xi_{1}
  & = &
        \frac{r_{1}}{r_{0}},
        \nonumber
  \\
  \xi_{2}
  & = &
        \frac{r_{2}}{r_{0}},
        \nonumber
  \\
  \xi_{M}
  & = &
        \frac{r_{M}}{r_{0}},
        \nonumber
  \\
  \vartheta_{0}
  & = &
        \frac{3V_{0}}{4\pi r_{0}^{3}},
\end{eqnarray}

\noindent
where $\vartheta_{0}$ is the ratio between the coordinate volume $V_{0}$
of the shell and the volume of an Euclidean sphere of radius $r_{0}$. The
expression in Equation~(\ref{Eqn46}) giving $r_{2}$ as a function of
$r_{1}$ is now translated as

\begin{equation}\label{Eqn49}
  \xi_{2}(\xi_{1})
  =
  \left(\vartheta_{0}+\xi_{1}^{3}\right)^{1/3}.
\end{equation}

\noindent
Note, for subsequent use, that this can also be written as
$\xi_{2}^{3}-\xi_{1}^{3}=\vartheta_{0}$. The relation which we must now
use in order to determine $\xi_{M}$ is that given in
Equation~(\ref{Eqn09}), which upon rescalings by $r_{0}$ can be written as

\begin{equation}\label{Eqn50}
  \sqrt{\frac{\xi_{2}}{3\left(\xi_{2}-\xi_{M}\right)}}
  =
  \sqrt
  {
    \frac
    {\xi_{1}}
    {\xi_{2}^{3}-\xi_{1}^{3}+3\left(\xi_{1}-\xi_{M}\right)}
  }
  +
  \frac{3}{2}
  \int_{\xi_{1}}^{\xi_{2}}d\xi\,
  \frac
  {\xi^{5/2}}
  {
    \left[
      \xi_{2}^{3}-\xi^{3}
      +
      3\left(\xi-\xi_{M}\right)
    \right]^{3/2}
  },
\end{equation}

\noindent
where we changed variables in the integral from $r$ to $\xi=r/r_{0}$.
Substituting for $\vartheta_{0}$ where possible we have the following
non-trivial algebraic equation that determines $\xi_{M}$ and therefore
$r_{M}$,

\begin{equation}\label{Eqn51}
  \sqrt{\frac{\xi_{1}}{\vartheta_{0}+3\left(\xi_{1}-\xi_{M}\right)}}
  -
  \sqrt{\frac{\xi_{2}}{3\left(\xi_{2}-\xi_{M}\right)}}
  +
  \frac{3}{2}
  \int_{\xi_{1}}^{\xi_{2}}d\xi\,
  \frac
  {\xi^{5/2}}
  {\left[\xi_{2}^{3}-\xi^{3}+3\left(\xi-\xi_{M}\right)\right]^{3/2}}
  =
  0.
\end{equation}

\noindent
Our objective here is to solve this equation in order to get
$\xi_{M}(\xi_{1})$, given a fixed value of $\vartheta_{0}$ and with
$\xi_{2}$ given by Equation~(\ref{Eqn49}). Note that, due to the
homogeneous scalings leading from the dimensionfull quantities to the
dimensionless ones, shown in Equation~(\ref{Eqn48}), each solution of this
equation is valid for any value of $\rho_{0}$, which no longer appears
explicitly. The same is true of the graphs to be generated using this
equation. Given a value of $\vartheta_{0}$, the corresponding graph
represents the results for all the possible strictly positive values of
the energy density $\rho_{0}$.

There are two main numerical tasks here, the calculation of the integral
and the resolution of this algebraic equation for $\xi_{M}$. The integral
can be readily and efficiently calculated by a cubic interpolation method,
using the values of the integrand and of its derivative at the two ends of
each integration interval. So long as we can return the value of the
integral without too much trouble, Equation~(\ref{Eqn51}) can be readily
and efficiently solved by an exponential sandwich (or bisection)
method~\cite{NumericalRecipes}. There are two readily available and robust
initial upper and lower bounds for the value of $\xi_{M}$, the minimum
possible lower bound being zero, and the maximum possible upper bound
being the energy of the unbound system, since we must have that
$Mc^{2}<M_{u}c^{2}$, which in terms of the dimensionless parameters
translates as $\xi_{M}<\vartheta_{0}/3$. We may therefore start the
process with a lower bound $\xi_{M\ominus}=0$ and an upper bound
$\xi_{M\oplus}=\vartheta_{0}/3$ for $\xi_{M}$. In practice, the efficiency
of this algorithm may be highly dependent on the use of a tighter pair of
bounds.

A few examples of the functions obtained in this way can be seen in
Figures~\ref{Fig02} through~\ref{Fig05}, which show $\xi_{M}$ as a
function of $\xi_{1}$, for fixed values of the energy of the unbound
system, that is, for fixed values of $\vartheta_{0}$. Each graph consists
of $81$ data points. In order to ensure good numerical precision we used
$10^{6}$ integration intervals in the domain $[\xi_{1},\xi_{2}]$. The
exponential sandwich was iterated until a relative precision of the order
of $10^{-12}$ was reached. The four graphs shown were generated on a
high-end PC in approximately $25$~hours, $15$~hours, $62$~hours and
$154$~hours, respectively, without too much preoccupation with efficiency.
As one can see, the graphs clearly display minima of $\xi_{M}$, which are
located at certain values of $\xi_{1}$. At these minima the pairs of
values $\left(\xi_{1},\xi_{2}\right)$ are given approximately, in each
case, by $(2.79,2.87)$, $(2.72,2.93)$, $(2.60,3.02)$ and $(2.35,3.21)$,
respectively. There is freely available an open-source
program~\cite{SSSLFPaperII} that can be used to perform these calculations
for any set of input parameters.

The minima of these functions give us the value of $\xi_{1}$ that
corresponds to the most tightly bound system that can be assembled from
the given unbound system in each case. With the given values of $\rho_{0}$
and $M_{u}c^{2}$, in each case this establishes the value of $r_{1}$ for
the most tightly bound and therefore energetically stable solution, and
hence determines the values of $r_{2}$, $r_{M}$ and of all the functions
describing both the spacetime geometry and the matter for that stable
solution. The limiting value of $\xi_{M}$ when $\xi_{1}\to 0$, not shown
in these graphs, corresponds to the interior Schwarzschild solution and
thus to the energy of the unbound system in each case, which in terms of
the variables shown in the graphs is given by $\vartheta_{0}/3$. The
$\xi_{1}\to\infty$ limit to the other side rises fairly slowly and does
not seem to approach this same value asymptotically, a situation that is
probably due to the fact that an infinitesimally thin shell at infinity
still has some binding energy, as compared to the corresponding set of
isolated infinitesimal point masses.

\section{Conclusions}\label{Sec05}

In this paper we have established the energetic interpretation of the
exact solutions obtained in a previous paper for spherically symmetric
shells of liquid fluid~\cite{LiquidShells}. All the energies involved were
precisely characterized, including the total energies of the unbound
systems, the total energies of the bound systems, the gravitational
binding energies, and the energies stored in the gravitational field. This
led to a characterization of the stability of the bound systems in terms
of their binding energies. We have identified a two-parameter family of
energetically stable solutions, within the original three-parameter family
of solutions. In a few cases the stable solutions were identified
numerically. It is to be expected that the interpretations of the energies
that were introduced here will be useful in other cases, such as those
involving polytropes, white dwarfs and neutron stars.

In order to accomplish this, integral expressions for all the energies
involved were presented, as integrals of the matter energy density over
various coordinate volumes. All these expressions hold more generally than
just in the case of constant energy density $\rho(r)=\rho_{0}$ that we are
directly dealing with here. A particular radial position $r_{z}$ within
the matter region, at which we have $\lambda(r_{z})=0$ and therefore
$\exp[\lambda(r_{z})]=1$ for the radial coefficient of the metric, was
identified as playing a special role in relation to the integral
expressions for the various energies. This is the single finite radial
position where the three-dimensional space is neither stretched nor
contracted, as compared to the behavior of the radial coordinate $r$.

The energetic interpretation was extended to the case of the two-parameter
family of interior Schwarzschild solutions for filled
spheres~\cite{SchwarzschildInternal,Wald}, which can be obtained as a
particular limit of the shell solutions, and which turn out to be
maximally unstable ones. This means that there is a strong tendency of the
solution for a filled sphere to spontaneously generate an internal vacuum
region and thus become a shell solution. This is clearly connected to the
repulsive character of the gravitational field around the origin, in the
case of the shell solutions, pushing matter and energy away from that
origin, as was discussed and characterized in the previous
paper~\cite{LiquidShells}. Any small perturbation of the interior
Schwarzschild solution will put this mechanism in action, thus leading to
an energetic decay from that filled sphere solution to a shell solution.

The crucial development leading to all this was the introduction of the
parameter $r_{\mu}$ in the previous paper, which was shown there to be
necessarily strictly positive in that case, for the correct resolution of
the differential equations and the corresponding interface boundary
conditions, as implied by the Einstein field equations. The apparently
traditional routine of choosing $r_{\mu}=0$ in order to eliminate the
singularity at the origin not only is often incompatible with the correct
resolution of the differential system but, when it is not thus
incompatible, it is tantamount to selecting a solution which has no
binding energy at all and is therefore maximally unstable from the
energetic point of view. Both from the purely mathematical point of view
and from the physical point of view, this is more often than not the
incorrect choice, which we are simply not at liberty to make.

\section*{Acknowledgments}

The author would like to thank his friends Prof. C. E. I. Carneiro and Mr.
Rodrigo de A. Orselli for their helpful criticism and careful reading of
the manuscript.

\bibliography{allrefs_en}\bibliographystyle{ieeetr}

\end{document}

%% file: packages.tex
\usepackage[latin1]{inputenc}
\usepackage{epsfig}
\usepackage{amsfonts}
\usepackage{cite}
\usepackage{color}
\definecolor{lgrey}{gray}{0.99}

%% file: math-defs.tex
\newcommand{\FFrac}[2]{{\displaystyle\frac{\displaystyle #1}{\displaystyle #2}}}

\newcommand{\e}[1]{\,{\rm e}^{#1}}